# The use of strain to tailor electronic thermoelectric transport properties: A first principles study of 2H-phase CuAlO$_2$

Evan Witkoske[1], David Guzman[4], Yining Feng[3], Alejandro Strachan[2], Mark Lundstrom[1], and Na Lu[2, 3]

[1]School of Electrical and Computer Engineering, Purdue University, West Lafayette IN, 47907
[2]School of Materials Engineering and Birck Nanotechnology Center, Purdue University, West Lafayette IN, 47907
[3]Lyles School of Civil Engineering, Purdue University, West Lafayette IN, 47907
[4]Condensed Matter Physics and Materials Science Department, Brookhaven National Laboratory, Upton NY, 11973

**Abstract:** Using first principles calculations, the use of strain to adjust electronic transport and the resultant thermoelectric (TE) properties is discussed using 2H phase CuAlO$_2$ as a test case. Transparent oxide materials, such as CuAlO$_2$, a p-type transparent conducting oxide (TCO), have recently been studied for high temperature thermoelectric power generators and coolers for waste heat. Given TCO materials with relative ease of fabrication, low cost of materials, and non-toxicity, the ability to tailor them to specific temperature ranges, power needs, and size requirements, through the use of strain opens an interesting avenue. We find that strain can have a significant effect on these properties, in some cases detrimental and in others beneficial, including the potential for n-type power factors larger than the highest p-type case. The physical reasons for this behavior are explained in the terms of the thermoelectric transport distribution and the Landauer distribution of modes.

## I. Introduction

Thermoelectric (TE) devices and materials hold great promise for broad use in solid-state energy generation and solid-state cooling. However, as robust and reliable as these devices are, they have been limited by low conversion efficiencies since their inception [1-5]. The past three decades have witnessed the thermoelectric material figure of merit, zT, improved from under one to over two [5]. These gains have been primarily driven by a reduction in the lattice thermal conductivity of materials and devices through the use of nano-structuring [6-12] and the development of novel materials that have an inherently low thermal conductivity due to large discrepancies in the masses of their constituent elements. These advances, however, have not translated into working devices [13]. As we approach the lower limit of the lattice thermal conductivity for common and even complex TE materials at room temperature and above, the variety of avenues capable of moving the field of thermoelectrics forward are being narrowed, therefore ideas that have the potential to advance the field need to be explored carefully.

In this paper we look at an alternate route forward, given materials with relative ease of fabrication, low cost, and non-toxicity, the ability to tailor them to specific temperature ranges, power needs, and size requirements through the use of strain opens an interesting avenue. Even though the overall zT efficiencies of these materials may not be able to beat state of the art TE materials, if the appropriate direction and magnitude of strain could be applied to increase their TE properties, the overall $cost/kW-hr of energy generation quite possibly could.

Because of their potential use in high temperature applications, due to a large band gap, high thermal stability, oxidation resistance, and low material costs, transparent conducting oxides (TCOs) have garnered interest for a variety of TE applications [14-25]. In this work 2H-phase $CuAlO_2$, which has gained interest as a promising candidate for high temperature p-type thermoelectric applications [15,18,26,27] due to the scarcity of p-type TCOs [28], under a variety of uniaxial, biaxial, and hydrostatic strains will be discussed. Limited theoretical and experimental studies have been done on the thermoelectric properties of the 2H phase of this material [15,28–30], and none to our knowledge have been conducted on the effect of strain on its thermoelectric properties. It will be shown that strain can have a significant effect on the band gap and electronic transport properties, in some cases detrimental and in others beneficial for thermoelectrics. This opens the door to the possibility of tuning the band gap as well as the electronic transport properties of TCOs to tailor them to specific thermoelectric applications [31–33].

There are five sections in this paper; I) introduction, II) atomic structure and methodology, III) electronic structure with and without strain, IV) thermoelectric transport properties of strained and unstrained structures, and finally, V) conclusions. We find that strain can offer both opportunities as well as challenges for thermoelectric device design with both the former and the latter being unique to the material and device required for specific applications.

## II. Atomic structure and methodology

$CuAlO_2$ crystallizes in two distinct phases, 3R and 2H, both having a delafossite structure with the rhombohedral (3R) and hexagonal (2H) phases occurring at atmospheric pressures [34]. In Fig. 1(a) the 2H phase structure, with a space group of $P6_3/mmc$ (no. 194), is shown with the crystallographic directions "a, b, and c" which are referred to as the [100], [010], and [001] directions throughout the paper. Figure 1(b) shows the high symmetry k-points of the first Brillouin zone, which are used for plotting the band structures.

All calculations were done using density function theory (DFT) as implemented in the open source package Quantum Espresso [35] to predict the atomic and electronic structure of $CuAlO_2$ under various strain conditions. The ab initio band structures (Kohn-Sham eigenvalues) were subsequently used to calculate general thermoelectric transport properties by utilizing the open source tool LanTrap 2.0 [36], which solves the Boltzmann Transport equation in the relaxation time approximation using the Landauer formalism [37]. The electron-ion interactions are accounted for using PAW, norm conserving pseudo-potentials; Al.pbe-n-

kjpaw_psl.0.1.UPF and Cu.pbe-dn-kjpaw_psl.0.2.UPF, along with the Ultrasoft, norm conserving pseudo-potential O.pbe-n-kjpaw_psl.0.1.UPF from the Quantum ESPRESSO pseudo-potential database. The electron exchange-correlation potential was calculated using the generalized gradient approximation (GGA)[38] within the Perdew-Burke-Ernzerhof (PBE) scheme. The kinetic energy cutoff for the expansion of the plane waves was set to 544.2 eV and all self-consistent calculations were terminated when a tolerance of 1.36 x 10$^{-5}$ eV in the total energy was reached. The structural relaxations were performed using a conjugate gradient (CG) algorithm and a 10x10x4 k-mesh. All structural relaxations were terminated when the force on all atoms are less than 2.57 meV/angstrom for the unstrained and hydrostatic (equal strain applied in all directions) cases, and 25.7 meV/angstrom for the uniaxial and biaxial strains. The electronic properties were computed on a finer 20x20x12 k-mesh.

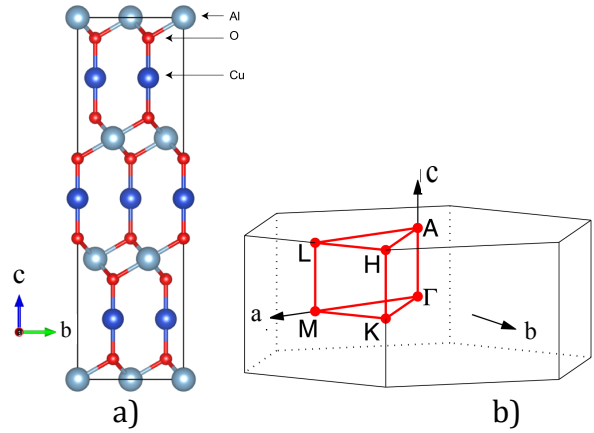

Figure 1. Fig. 1(a) is the relaxed conventional super cell of 2H-phase $CuAlO_2$. Fig. 1(b) is the first Brillouin zone with the high symmetry points used for the dispersion paths shown in Figs. 2-5.

## III. Electronic structure with and without strain

### A. Relaxed band structure

The lattice constants in the relaxed structure were found to be a = b = 2.855 Å and c = 11.394 Å which agree well with experimental [39] and theoretical [29,39] results. Band gap calculations using DFT, a ground state method, generally do not produce reliable results due to the excited-state nature of the band gap, as well as derivative discontinuities in the exchange-correlation energy functional [40,41] arising when the number of electrons increases by an integer step at the transition between the highest occupied and lowest unoccupied single electron level in an N-electron system [42]. Notably however, the theoretical band gap prediction in the current work as well as from other groups shows remarkable accuracy to experimental results. Our calculations give an unadjusted indirect band gap of 1.85 eV, which is similar to the experimental results of 1.8 eV by Yanagi [43] and 1.65 eV by Benko [44], as well as theoretical values of 1.85 eV by Jayalakshmi [29], 1.82 eV by Liu [45], and 1.7 eV by Yanagi [43]. The detailed explanation for this theoretical accuracy won't be discussed.

Figure 2(a) shows the conduction band minimum occurs at the Γ point while the valence band maximums are located at M, K, L, and H, all with similar energies. The valence band of 2H $CuAlO_2$ is built from the hybridization between the oxygen 2p and the aluminum 3s and 3p states in addition to the copper 3d states, which contribute the majority of states at the valence band edge. Copper 4s, aluminum 3s, and aluminum 3p states mainly form the conduction band. The interesting valence band structure at all four valence band maximum points should be noted and will be discussed in Sec. IV.

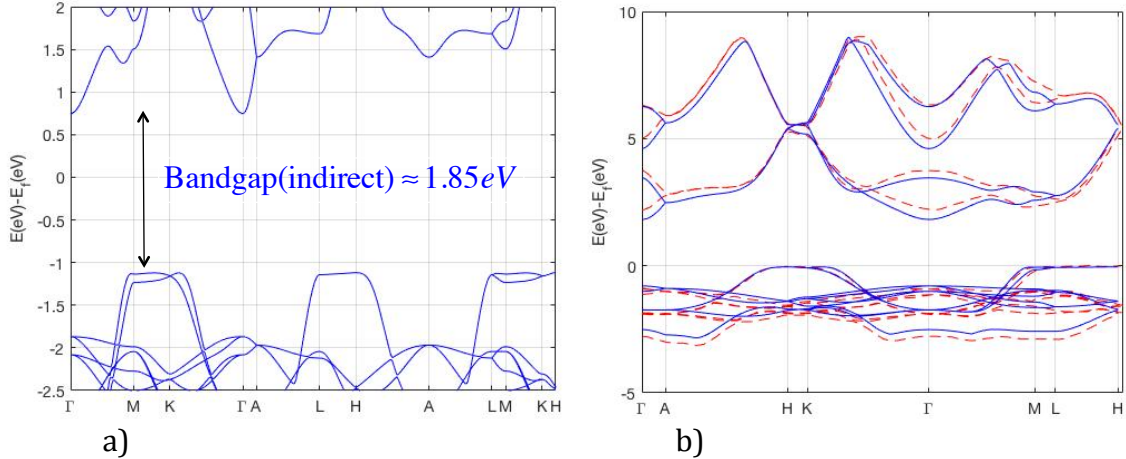

a)  b)

Figure 2. Fig. 2(a) is the calculated band structure of the 2H phase of $CuAlO_2$. Fig. (b) is a comparison of our electronic band structure (solid blue lines) to that of reference [30] (dashed red lines), (note figures 2(a) and 2(b) use different k-paths). For all band structures in this paper, the Fermi energy is normalized to zero when calculating band alignment, to allow visual comparisons with a variety of strained and relaxed band structures.

**B. Band structures with hydrostatic strain**

DFT calculations are done at 0K, so for materials that are useful at higher temperatures, such as $CuAlO_2$, once the unstrained structures' band gap is confirmed, adding negative hydrostatic stress can be a useful guide to help find trends in the band gap at these elevated temperatures due to the effects of thermal strain. Imparting confidence in this methodology, the lattice parameters under hydrostatic strain for this study are found to be consistent with theory and experimental values [46,47].

To simulate hydrostatic strain (equal strains in all directions), we took the relaxed structure and applied isotropic strain to the cell parameters by plus/minus 1,2, and 3%. The atomic positions were then allowed to relax keeping the volume of the cell constant. In all the strained cases shown in Fig. 3, most of the difference is a small band gap and electron affinity adjustment. An outlier is the -3% strained case, in which we see a drastic electron affinity adjustment. Most of the shape of the unstrained band structure remains when strain is applied in all six cases studied.

The conduction band minimums and valence band maximums remain at the same high symmetry k-points as for the unstrained case. There are some variations in the curvature of the band structure much higher in the conduction band for the 3% compressive and tensile cases. It should be noted that there is very little, if at all, change in the structure of the valence band near the band edge, which is the most relevant area to p-type TE transport properties. A summary of the changes in band gaps can be found in Table 1.

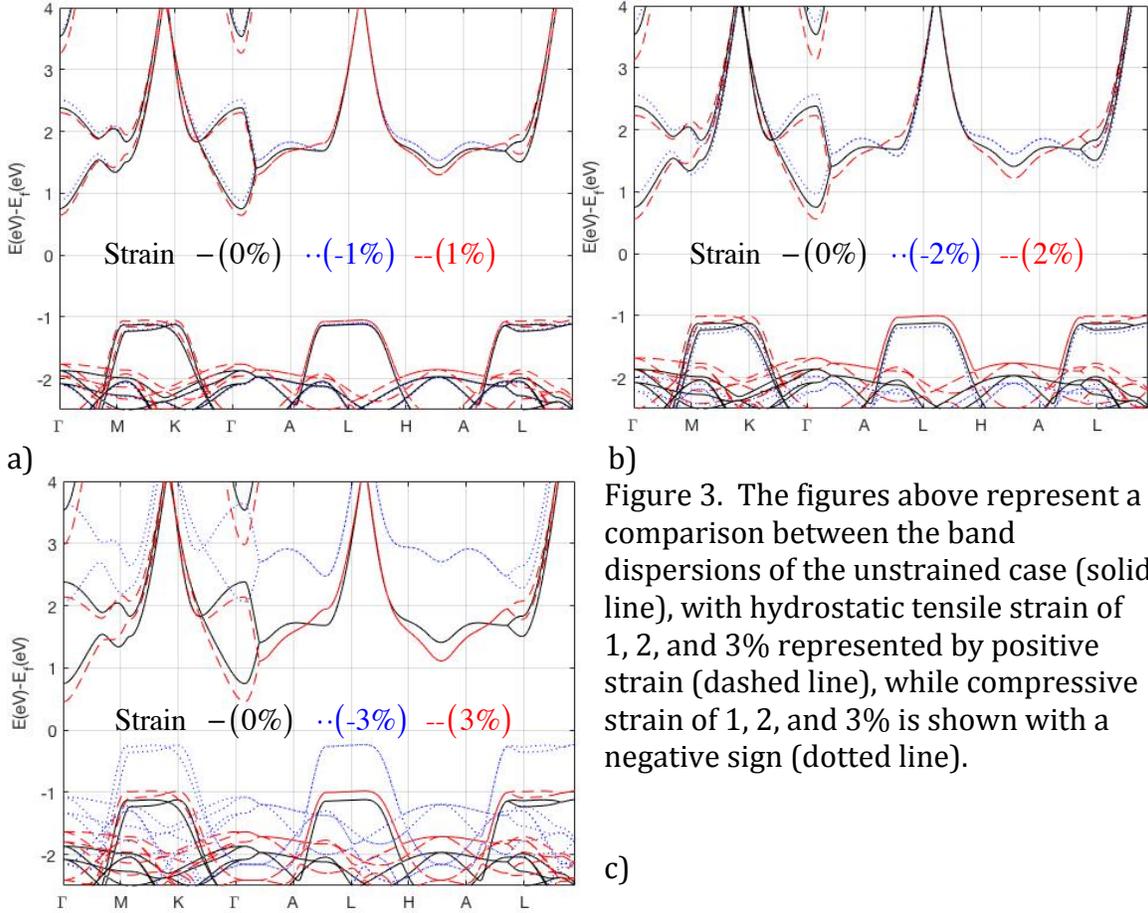

Figure 3. The figures above represent a comparison between the band dispersions of the unstrained case (solid line), with hydrostatic tensile strain of 1, 2, and 3% represented by positive strain (dashed line), while compressive strain of 1, 2, and 3% is shown with a negative sign (dotted line).

## C. Uniaxial strain

We applied strain of ±1% in the [100], [010], or [001] crystallographic direction and allowed the other transverse directions to relax. There are two main observations from the band dispersions in Fig. 4. As compared to the unstrained case in Fig. 2, the valence and conduction bands have changed substantially in all the structures simulated in Fig. 4. The strained structures have lost the flat valence band and have been replaced by a more parabolic band dispersion in four of the six cases (-1% [100], ±1% [010], and -1% [001]). This will have an effect on electron transport that will be quantified in Sec. IV. The second immediate observation is the changes in band gap and band gap minimum locations in the above structures. The band gap has disappeared for the -1% [100] and [001] strains as well as both ±1% [010] cases, and reduced for the 1% [100] and [001] cases. The significant reduction in band gap will hurt thermoelectric performance at high temperatures due to bipolar effects. The conduction band minimum remains around the Γ point, however a second indirect band gap minimum has been reduced in energy and lies at a different high symmetry point for many of the cases in Fig. 4.

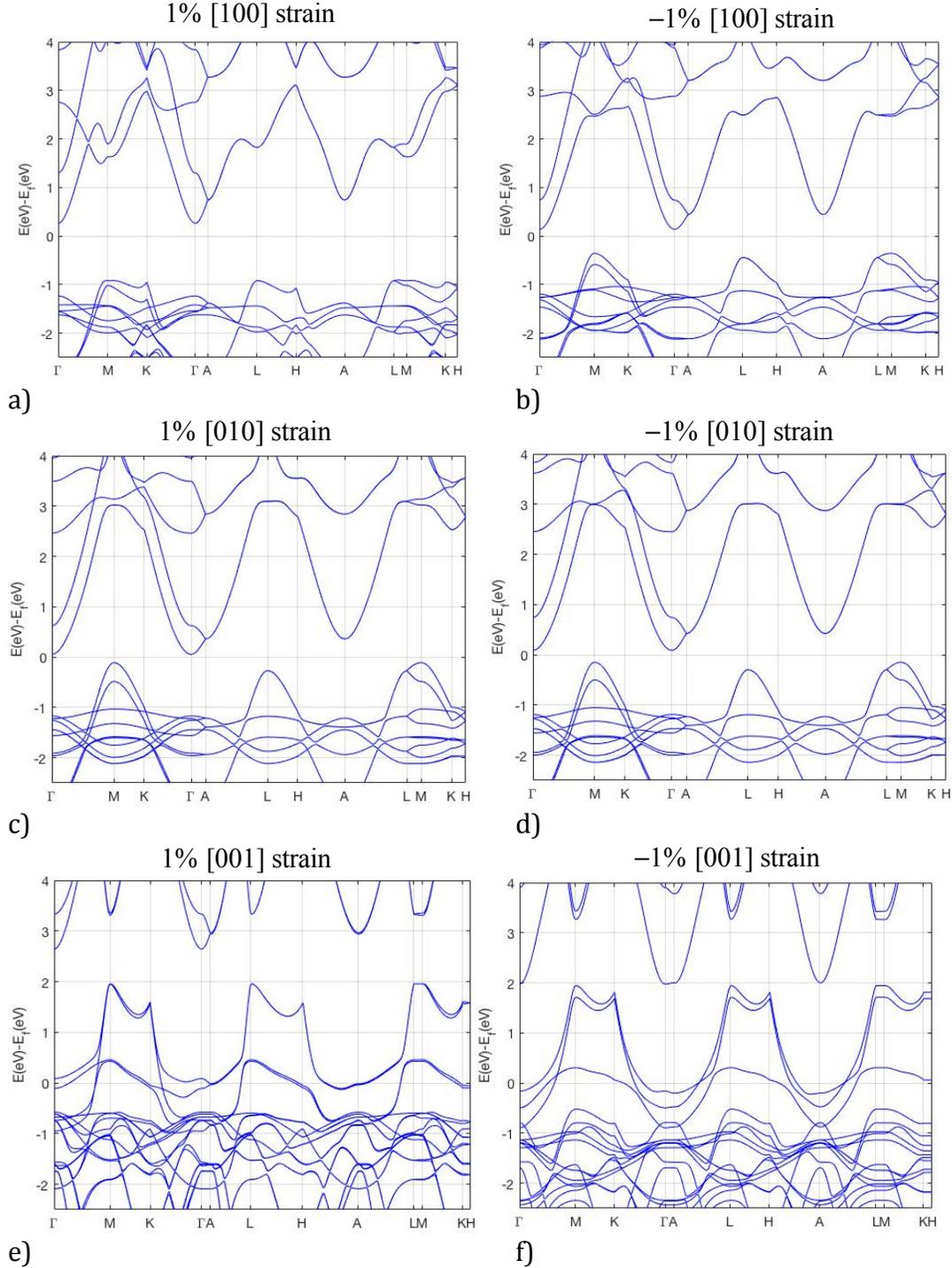

Figure 4. The band dispersion relation of 2H-CuAlO$_2$ with ±1% uniaxial strain applied to the [100], [010], and [001] directions respectively, with the other two directions allowed to relax.

## D. Biaxial strain

For biaxial strain the same procedure was used as section C), except we apply ± 1% strain to two of the three directions and allowed the third direction to relax. Figure 5 is the band dispersion relation for ± 1% strain applied to the [110] direction, as well as the +1% [101] strain case. For the other structures computed, (i.e. ± 1% [011] and -1% [101]), the band gaps disappear entirely and the material becomes metallic for the energy range of interest. With no band gap, the Fermi level lies within the bands, most of the heat is carried by electrons and holes, therefore the overall Seebeck coefficient will be close to zero, due to their combined, offsetting effects. We looked only at structures that retain their semiconducting properties.

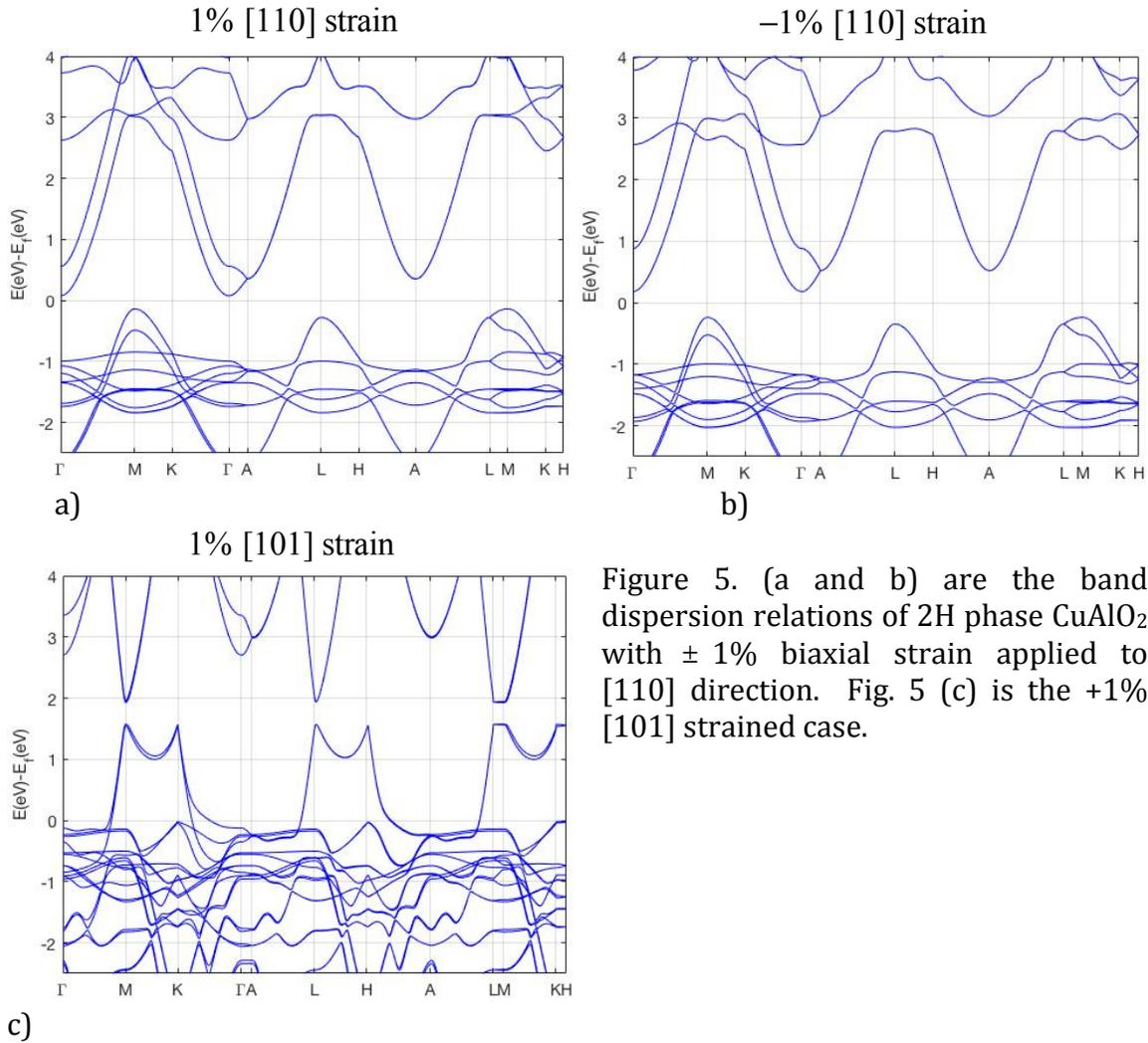

Figure 5. (a and b) are the band dispersion relations of 2H phase $CuAlO_2$ with ± 1% biaxial strain applied to [110] direction. Fig. 5 (c) is the +1% [101] strained case.

The main observations from the band dispersion in Fig. 5 are similar to those from the uniaxial strain case in Fig. 4. The valence band has changed substantially in structure compared to Figure 2(a). The interesting structure of unstrained $CuAlO_2$, having a very flat valence band, has disappeared and been replaced by a more traditional band dispersion in the [110] strain cases. The second immediate observation is that much of the band gap has disappeared for all three cases in Figure 5. The +1% [101] strain case has changed from an indirect band gap to direct, while retaining a secondary indirect band gap of similar energy. The valence and conduction bands have become very sharp (i.e. small effective mass) in the +1% [101] strain case. All cases are summarized in Table 1 below.

**Table 1)**

| Strain Type | | $E_{gap}$ (eV) | | 2$^{nd}$ $E_{gap}$ (within 0.5 eV of primary $E_{gap}$) | | |
|---|---|---|---|---|---|---|
| Unstrained | (I) | 1.85 | Γ→M & Γ→L | None | | |
| | | | | | | |
| Hydrostatic -3% | (I) | 2.33 | Γ→M & Γ→L | None | | |
| -2% | (I) | 2.16 | Γ→M & Γ→L | | | |
| -1% | (I) | 2.00 | Γ→M & Γ→L | | | |
| +1% | (I) | 1.72 | Γ→M & Γ→L | None | | |
| +2% | (I) | 1.57 | Γ→M & Γ→L | | | |
| +1% | (I) | 1.44 | Γ→M & Γ→L | | | |
| | | | | | | |
| [110]  -1% | (I) | 0.41 | Γ→M | (I) | 0.52 | Γ→L |
| +1% | (I) | 0.20 | Γ→M | (I) | 0.34 | Γ→L |
| | | | | | | |
| [101]  +1% | (D) | 0.35 | M→L | (I) | 0.38 | M→K & L→H |
| | | | | | | |
| [100]  -1% | (I) | 0.51 | Γ→M | (I) | 0.57 | Γ→L |
| +1% | (I) | 1.18 | Γ→M & Γ→K | (I) | 1.65 | A→L & A→M & A→K |
| | | | | | | |
| [010]  -1% | (I) | 0.24 | Γ→M | (I) | 0.39 | Γ→L |
| +1% | (I) | 0.16 | Γ→M | (I) | 0.33 | Γ→L |
| | | | | | | |
| [001]  -1% | (I) | 0.02 | Γ→M & Γ→L | (I) | 0.05 | A→L |
| +1% | (I) | 0.69 | Γ→M | (I) | 1.05 | Γ→K    (I) 0.97 L→M |

Table 1). Summary of the corresponding strain dimension and % applied is followed by the band gap energy. The (I) (indirect) and (D) (direct) is followed by the corresponding symmetry point(s) where the gap minimums in energy occur.

### IV. Thermoelectric Transport

**A. Landauer transport method**

The performance of a thermoelectric material is directly related to its material figure of merit,

$$zT = \frac{S^2 \sigma T}{\kappa_L + \kappa_e}, \tag{1}$$

where $S$ is the Seebeck coefficient, $\sigma$ the electrical conductivity, $\kappa_L$ and $\kappa_e$ the lattice and electronic thermal conductivities, and $T$ is the temperature. The thermoelectric transport parameters

$$\sigma = \int_{-\infty}^{+\infty} \sigma'(E) dE \tag{2a}$$

$$S = -\frac{1}{qT} \int_{-\infty}^{+\infty} (E - E_F) \sigma'(E) dE \bigg/ \int_{-\infty}^{+\infty} \sigma'(E) dE \tag{2b}$$

$$\kappa_0 = \frac{1}{q^2 T} \int_{-\infty}^{+\infty} (E - E_F)^2 \sigma'(E) dE = \kappa_e + T\sigma S^2 \tag{2c}$$

with the differential conductivity, $\sigma'(E)$, given by

$$\sigma'(E) = q^2 \Xi(E)(-\partial f_0 / \partial E), \tag{2d}$$

and the transport distribution in the diffusive limit written in the Landauer form [48],

$$\Xi(E) = \frac{2}{h}(M(E)/A)\lambda(E), \tag{2e}$$

with $M(E)/A$ being the number of channels per cross-sectional area for conduction and $\lambda(E)$ being the mean-free-path (MFP) for backscattering (See the appendix in [49] for a short derivation of (2e) and [37] for a longer discussion).

In (2e), the mean-free-path for backscattering is defined as [37]

$$\lambda(E) \equiv 2v_x^2(E)\tau_m(E)/v_x^+(E), \tag{3a}$$

where $v_x^2(E)$ is an average over angle of the quantity $v_x^2(\vec{k})$ at energy, $E$. The velocity, $v_x^+(E)$, is the angle-averaged velocity in the +x direction (See [37] for the definitions of these averages). The number of channels at energy, $E$, is [37,50]

$$M(E)/A = hv_x^+(E)D(E)/4, \tag{3b}$$

where $D(E)$ is the density-of-states per unit volume including a factor of two for spin.

The numerical methods used to calculate eqns. (1) and (2) in LanTrap [36], using a band structure from density functional theory (DFT) simulations as input, are described in the supplementary information of [49,51]. In this work we do not consider the variability of the lattice thermal conductivity, $\kappa_L$, with strain [31,46,47,52]. Therefore, if we assume a constant lattice thermal conductivity and a small electronic thermal conductivity, $\kappa_e$, due to a relatively low electrical conductivity for oxides, for our purposes in this work, we will focus on the power factor given by

$$PF = S^2\sigma, \qquad (4)$$

which is comprised of the Seebeck coefficient and the electrical conductivity.

### B. Scattering

When calculating eqns. (1), (2), and (4), a constant mean free path (MFP) of 3 nm for both holes and electrons was used. This is consistent with the small nano-scale size limit (SNS) [53,54] due to grain boundary scattering [55] of nano-structured TE materials, which has been used to reduce the overall lattice thermal conductivity. In general, to get a more complete understanding of the scattering mechanisms involved in a particular material or device, experimental mobility should be measured for the structure of interest, or first principles guided simulations performed [49], to help elucidate the type of scattering mechanisms and their coupling strength.

Having assumed a constant mean-free-path of 3 nm, the transport distribution of eq. (2e) now depends only on the distribution of modes eqn. (3b). This greatly simplifies comparing different band structures to ascertain which will provide the largest PF or zT, the only quantity needed in this case is $M(E)$, which varies proportionally with the density of states. Therefore, in the case of a constant MFP, an increase in the density of states right around the valence band edge will increase the distribution of modes, thereby increasing the power factor. In this case, a large density of states at the valence band edge is very beneficial [56], however if the scattering rate were taken to be acoustic deformation potential (ADP) scattering, which goes inversely with density of states, the benefits aren't always clear [49].

### C. Discussion

Figure 6 is a comparison of the transport properties in the [100], [010], and [001] directions of transport for the unstrained case assuming a MFP of 3 nm at 300 K. In all plots the x-axis is the Fermi level with the valence band located at 0 eV and the conduction band located at 1.85 eV. The largest power factor, Fig. 6(a), is obtained with transport in the [001] direction in the unstrained structure, this is also true for all strain cases considered next.

Fig. 6(b) is a plot of the conductivity vs. Fermi level for the three different transport directions considered. The values of the conductivity show in Fig. 6(b) and Table 2 at the valence band edge are similar to those reported theoretically using a constant scattering time [30]. However, these conductivity values are approximately 1-2 orders of magnitude larger than what have been reported experimentally [30,57,58]. This is attributed to the 1-2 orders of magnitude lower carrier concentrations in experiment, i.e. $\sim 1\times 10^{18}$ to $1\times 10^{19}$ cm$^{-3}$ compared to theoretical carrier concentrations at the power factor maximizing Fermi level around the valence band edge, i.e. $\sim 1\times 10^{20}$ to $1\times 10^{21}$ cm$^{-3}$.

Fig. 6(c) is a plot of the Seebeck coefficient vs. Fermi level. A materials' maximum Seebeck coefficient is not dependent on the overall effective mass directly, but only on band gap and weakly on the scattering mechanism (i.e. Ionized Impurity, Acoustic Deformation Potential Scattering (ADP), constant MFP, etc.). The overall Seebeck vs. Fermi level curve will be unchanged for the three different transport directions considered. The values shown are consistent with other theoretical studies [30] which comes as no surprise since Seebeck values are directly related to the band gap.

Figure 6(e) is a plot of the distribution of modes for the three transport directions. The band structure with the largest $M(E)$ at the band edge will also have the largest power factor, which can be seen in Fig. 6(a). We notice in Fig. 6(e) the dramatically different distribution of modes around the valence and conduction band edges. For parabolic band semiconductors in three dimensions, $M(E)$ varies linearly with energy around the conduction and valence band edges as $M(E) \sim |(E_{C,V} - E)|$ [37,50]. Focusing on the [001] direction, the distribution of modes at the conduction band edge goes linearly with energy as expected. The distribution of modes at the valence band however has a strikingly non-linear shape, leading to the larger power factor values obtain for p-type as compared to n-type carriers in the unstrained case.

We have summarized the power factor, Seebeck coefficient, and electrical conductivity at the Fermi level that maximizes the power factor for all of the strained and unstrained cases in Table 2. In all strained and unstrained cases, transport in the [001] direction yields the highest power factor. The Fermi level that maximizes the power factor can vary depending on band structure and scattering, however in all cases discussed in this work, the Fermi level that maximizes the power factor lies very close to the valence band edge, as can be seen from Fig. 6(a).

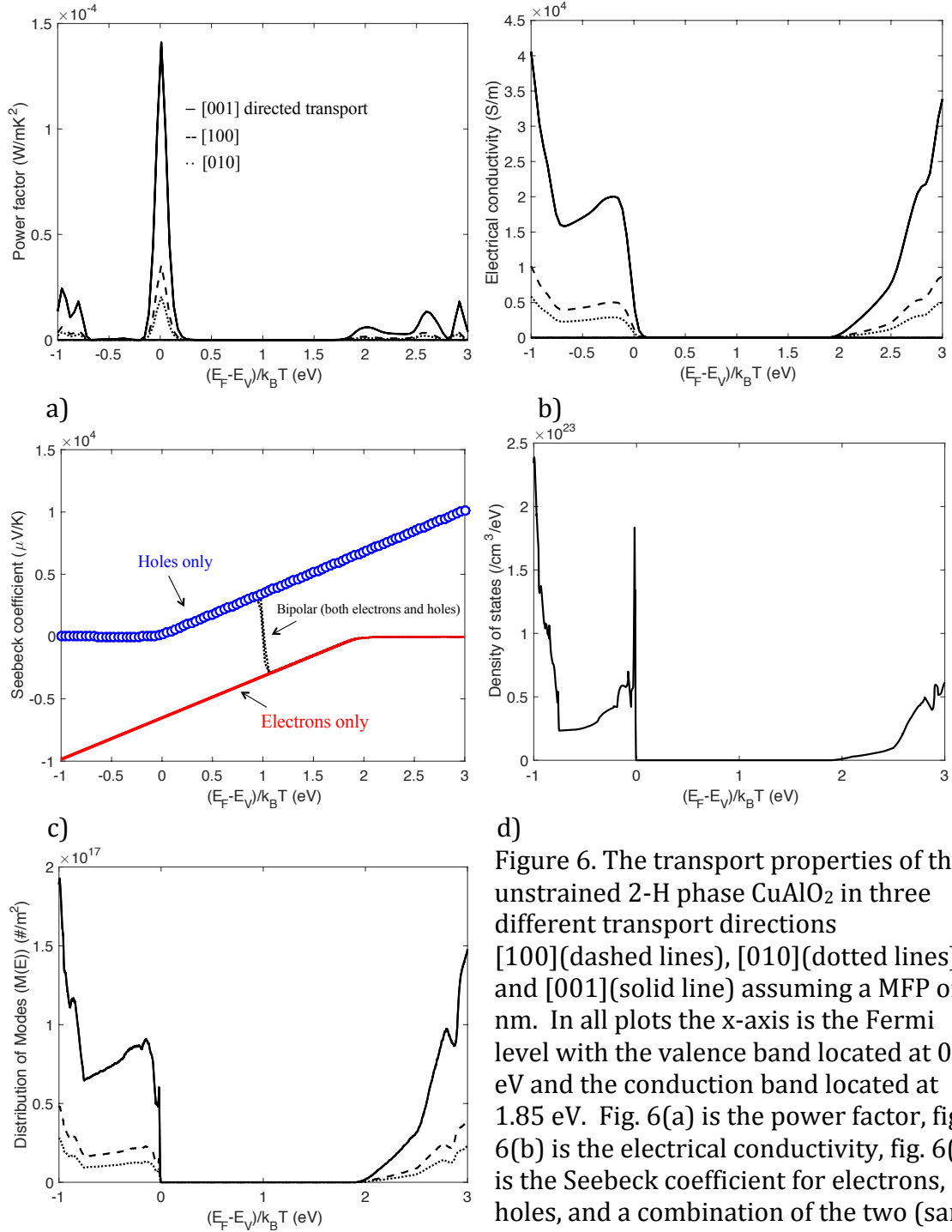

Figure 6. The transport properties of the unstrained 2-H phase $CuAlO_2$ in three different transport directions [100](dashed lines), [010](dotted lines), and [001](solid line) assuming a MFP of 3 nm. In all plots the x-axis is the Fermi level with the valence band located at 0 eV and the conduction band located at 1.85 eV. Fig. 6(a) is the power factor, fig. 6(b) is the electrical conductivity, fig. 6(c) is the Seebeck coefficient for electrons, holes, and a combination of the two (same for all transport directions), fig. 6(d) is the DOS (same for all transport directions), fig. 6(e) is the distribution of modes.

In Table 2, the hydrostatic strain cases' power factors closely resemble the unstrained case, which should be expected. When the band gap is adjusted due to hydrostatic pressure, the Seebeck coefficient value at a given energy and k point changes. The Fermi level is adjusted to maximize the power factor, changing the value of the Seebeck coefficient and conductivity at this new Fermi level [49]. Due to these adjustments, the power factors end up being about the same as the unstrained case.

The two cases that raise interest in Table 2 due to their higher power factors are the +1% [101] and the +1% [001] strains. Both have a higher conductivity at the valence band edge than any other cases. The +1% [001] strain yields a p-type power factor of $1.95 \times 10^{-4}$ (W/mK$^2$), the highest p-type value of the strains considered. The shape of the valence band for the +1% [101] and +1% [001] strained cases are very similar, with the effective masses becoming very small due to the large curvature of the valence band at the gamma point as can be seen from Figs. 4 and 5(c). These small effective masses, along with the addition of a second indirect peak at the K and H high symmetry k-points, facilitate a higher distribution of modes, due to the increase of the density of states with the addition of a second band, and a higher positive directed velocity (smaller effective mass), both of which contribute to an increase in $M(E)$. In both of these high power factor cases, n-type conduction has an even higher power factor than p-type. This is due to the same effects described for the increase in p-type conduction. Note the -1% [001] case also has very high conductivity values, however with a negligible band gap, both holes and electrons contribute to the conductivity, making the overall power factor low. Due to the band gaps of 0.35 eV for +1% [101] and 0.69 eV for +1% [001], the power factors for both n and p-types remain high.

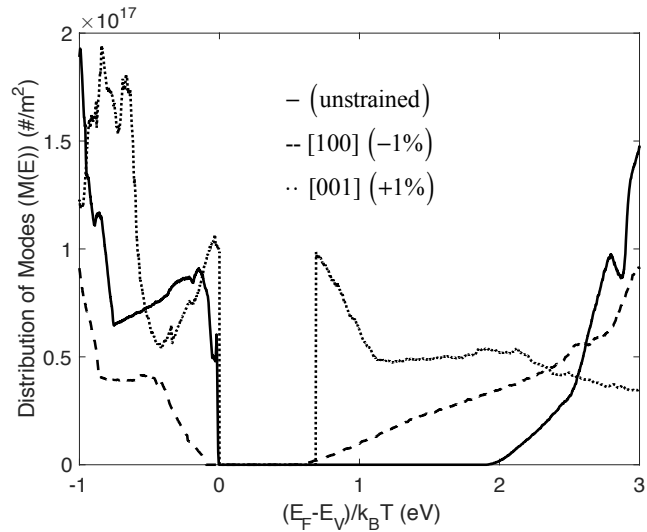

Figure 7. Distribution of modes for three cases; the unstrained case (solid line), the -1% [100](dashed line) and the +1% [001](dotted line) strained cases. Because of the constant MFP assumption, $M(E)$ is also the transport distribution, $\Xi(E)$.

Figure 7 is a plot of the distribution of modes for three different strained cases; unstrained, -1% [100], and +1% [001], which correspond to our unstrained reference, the lowest, and the highest power factor cases for comparison. As was mentioned before, when the same constant MFP is assumed for all cases, eqns. (2d-2e) show that only the Fermi window and distribution of modes determine the conductivity. Therefore, the structure with the largest distribution of modes will also have the largest power factor.

In Fig. 7, the valence band edges of all three cases are shown at 0 eV on the x-axis, but due the strain applied, the conduction band edges lie at 1.85, 0.51, and 0.69 eV respectively. The abrupt n-type distribution of modes for +1% [001] at the conduction band can also be seen in Fig. 7. This illustrates the direct correlation between the maximum power factor and the maximum distribution of modes at the band edge.

**Table 2)**

| Strain Type | | p(n)-type electrical conductivity x $10^3$ (S/m) (@$E_F$=$E_{PF(max)}$) | Seebeck coefficient ($\mu$V/K) (@$E_F$=$E_{PF(max)}$) | Max Power Factor x $10^{-4}$ (W/mK$^2$) (Optimum transport direction) |
|---|---|---|---|---|
| Unstrained | | 3.82 | 192 | 1.41 |
| Hydrostatic | -3% | 3.36 | 207 | 1.44 |
| | -2% | 3.55 | 201 | 1.43 |
| | -1% | 3.73 | 196 | 1.43 |
| | +1% | 3.90 | 191 | 1.42 |
| | +2% | 3.94 | 190 | 1.42 |
| | +3% | 3.92 | 190 | 1.41 |
| [110] | -1% | 0.75 | 197 | 0.29 |
| | +1% | 0.95 | 169 | 0.27 |
| [101] | +1% | (p) 7.70  (n) 6.81 | (p) 145  (n) 161 | (p) 1.63  (n) 1.77 |
| [100] | -1% | 0.79 | 142 | 0.16 |
| | +1% | 0.78 | 168 | 0.22 |
| [010] | -1% | 0.62 | 213 | 0.28 |
| | +1% | 0.66 | 210 | 0.29 |
| [001] | -1% | (p) 10.2  (n) 10.3 | (p) 17.1  (n) 13.9 | (p) 0.03  (n) 0.02 |
| | +1% | (p) 2.89  (n) 4.75 | (p) 260  (n) 206 | (p) 1.95  (n) 2.02 |

Table 2) Summary of the electrical conductivity, Seebeck coefficient, and power factor at the Fermi level that maximizes the power factor. All calculations were done in LanTrap 2.0 [36] with a MFP of 3 nm with bipolar effects included.

## V. Conclusion

Due to the lack of p-type thermoelectric materials for high temperatures, materials that offer interesting band structure warrant careful consideration. In this work, we used first principles calculations to analyze a promising p-type thermoelectric material that is earth abundant, robust at high temperatures, and oxidation resistant. The drastic change in structure produced when strain is applied creates a type of band dispersion that needs careful analysis to ascertain the benefits and detriments for high temperature (and possibly low temperature) thermoelectrics. All TE transport parameters are determined by the transport function. In the SNS scattering limit assumed in this study (i.e. a constant MFP), the transport function is proportional to the number of channels, $M(E)$. Therefore, the effect of strain on TE transport is best understood by examining the $M(E)$ extracted from the bandstructure. A large number of channels near the band edge lead to a high electrical conductivity.

The main conclusions of this study are: 1) strain can be beneficial or detrimental to TE performance depending on whether it increases or decreases $M(E)$ near the band edge, and 2) under certain cases of strain, n-type conduction produced higher power factors than their p-type counterparts, thus opening an interesting avenue for strain engineering to produce both n and p type legs from the same material. The enhanced n-type performance occurs because the right type of strain dramatically increases $M(E)$ near the conduction band edge. These results suggest great care must be undergone in the fabrication of this material to prevent detrimental strains, which can lead to degradation of thermoelectric performance. Conversely however, there are also benefits if care is undertaken in fabrication to produce thermoelectric materials that outperform their unstrained cases.

Many researchers feel that thermoelectrics could potentially provide a robust source of energy for a rapidly growing and energy consuming population. Transparent conductive oxide (TCO) materials are attractive because they offer relative ease of fabrication, low cost of materials, and non-toxicity. The ability to tailor TCO materials to specific temperature ranges, power needs, and size requirements, through the use of strain would open up interesting new avenues. Although the overall zT efficiencies of TCO materials may not exceed state of the art TE materials, if the appropriate direction and magnitude of strain could be applied to increase their TE properties, the overall $cost/kW-hr of TCO's quite possibly could.

**Acknowledgements** – This work was partially supported by NSF CAREER project (CMMI 1560834)

*Data Availability* – Access to scripts and the computational tool LanTrap used in this work are be available for free online at nanohub.org/groups/needs/lantrap. The pseudo-potentials used in the DFT calculations can be found at http://www.quantum-espresso.org/pseudopotentials.